\documentclass[aps,prl,showpacs,preprint,groupedaddress]{revtex4}
\usepackage{graphicx}

\begin{document}

\setlength{\belowcaptionskip}{5pt}
\setlength{\abovecaptionskip}{5pt}
\setlength{\textfloatsep}{5pt}

\title{Negative Poisson's ratio materials via isotropic interactions}
\author{Mikael C. Rechtsman$^1$}
\author{Frank H. Stillinger$^2$}
\author{Salvatore Torquato$^{2,3,4,5}$} 
\affiliation{$^1$Department of Physics, Princeton University,
Princeton, New Jersey, 08544} 
\affiliation{$^2$Department of Chemistry, Princeton University,
Princeton, New Jersey, 08544} 
\affiliation{$^3$Program in Applied and
Computational Mathematics and PRISM, Princeton, New Jersey, 08544}
\affiliation{$^4$Princeton Center for Theoretical Physics, Princeton,
New Jersey, 08544} 
\affiliation{$^5$School of Natural Sciences, Institute for Advanced Study, Princeton, New
Jersey, 08544}
 \date{\today}
\begin{abstract}
We show that under tension, a classical many-body system with only isotropic pair interactions in a crystalline state can, counterintutively, have a negative Poisson's ratio, or auxetic behavior.  We derive the conditions under which the triangular lattice in two dimensions and lattices with cubic symmetry in three dimensions exhibit a negative Poisson's ratio.  In the former case, the simple Lennard-Jones potential can give rise to auxetic behavior.  In the latter case, negative Poisson's ratio can be exhibited even when the material is constrained to be elastically isotropic.
\end{abstract}
\maketitle

Materials with negative Poisson's ratio (NPR), the so-called ``auxetics," are those that 
when stretched in a particular direction, expand in an orthogonal direction.  NPR 
behavior is a counterintuitive material property that has been observed only in a handful 
of materials that often have intricate structures and characteristic lengths much larger 
than an atomic bond length.  NPR materials have a great deal of technological potential, for example, to increase the sensitivity of piezoelectric transducers \cite{paper149}; as components in microelectromechanical systems; and as shock absorbers and fasteners \cite{TorquatoWhitesidesNPR, ChoiFasteners}.

A negative Poisson's ratio had not been observed in any elastically isotropic material until 
the discovery of certain foam structures with
re-entrant structures \cite{Lakes1987}.  NPR behavior has also been observed in multiple-length-scale laminate composites 
\cite{MiltonComposites}, polymeric and metallic foams \cite{MetallicPolymericFoams}, 
inverted honeycomb and other structures fabricated using soft lithography 
\cite{TorquatoWhitesidesNPR}, and scaffold structures made out of 
springs, hinges, and rods \cite{AlmgrenHinges}.  It has been found in cubic atomic solids when they are stretched in the [110] direction \cite{BaughmanNPR}.    

In this Letter, we derive conditions for which NPR behavior is exhibited in 
classical many-body systems; this continues a research program in 
which inter-particle interactions are sought for targeted material properties.  Examples of such \emph{inverse problems} include optimization of pair potentials to give rise to negative thermal expansion 
\cite{RechtsmanNTE}, and a method to derive potentials that yield targeted classical 
ground states \cite{RechtsmanPRL}.

We report here that under tension, two- and three-dimensional systems with 
isotropic two-body interaction potentials can show NPR behavior in the crystal 
phase as long as certain linear equalities and inequalities involving the interaction 
potential are satisfied.  This is an unexpected result, since an inherently 
anisotropic behavior arises from isotropic interactions; indeed, most previously 
discovered NPR materials exhibit complex, carefully designed anisotropic interactions.  
We show this to be the case at zero temperature for the elastically isotropic triangular 
lattice in two dimensions, and for the fcc lattice in three dimensions.  In the latter case, 
NPR behavior is exhibited even when the material is constrained to be elastically isotropic.  We first describe the calculation of the Poisson's ratio for any dimension.  Then, we present results for the two- and 
three-dimensional cases, including the elastic constants and NPR constraints.  In order to demonstrate that NPR behavior is achievable at positive pressure, we present an example 
in which this is achieved by including three-body interactions in two or three dimensions.

Consider a set of $N+1$ particles, with positions $\left\{{\bf r}_n\right\}$ ($n\geq 0$), that 
occupy a particular Bravais lattice in a state of zero strain.  Under strain that is uniform 
throughout space, the new positions of the particles are
\begin{equation}
{\bf x}_n = \left({\bf I}+{\bf E}\right)\cdot {\bf r}_n,
\end{equation}
where ${\bf I}$ is the unit tensor in $d$ dimensions and ${\bf E}$ is the second-rank strain 
tensor with components $\varepsilon_{ij}$.  The latter is constrained to be symmetric (i.e., 
$\varepsilon_{ij}=\varepsilon_{ji}$) in order to remove simple rotation.  For simplicity, we 
take the origin to be at ${\bf r}_0$, and thus the energy per particle of the system can be 
written as
\begin{equation}
u = \frac{1}{2}\sum_{n=1}^N \phi(|{\bf x}_n|),\label{energy}
\end{equation}
where $\phi$ is the pair interaction potential and $N$ is the total number of particles 
excluding that at the origin.  At zero temperature, the enthalpy and Gibbs free energy per 
particle are equivalent and equal to $g = u + pv$, where $p=-du/dv$, and $v$ is the 
$d$-dimensional specific volume.  The equilibrium volume is found by minimizing $g$ 
with respect to $v$ at fixed $p$.  In order to calculate the second-order elastic constants 
and Poisson's ratio, $g$ is expanded to quadratic order in the strain tensor, 
$\varepsilon_{ij}$.  The expansion is taken around equilibrium; thus, it 
contains no linear terms, and can be written as $g \simeq g_0+\frac{v}{2}\lambda_{ijkl}
\varepsilon_{ij}\varepsilon_{kl}$, where $g_0$ is the zero-strain free energy, $
\lambda_{ijkl}$ are the second-order elastic constants of the system \cite{LandauLifshitz}, 
and the Einstein convention is used.  The Poisson's ratio $\nu=-
\varepsilon_T/\varepsilon_L$ is calculated by imposing a strain, $
\varepsilon_L=\varepsilon_{ij}n_in_j$, and minimizing the free energy to find the 
strain in a transverse direction, $\varepsilon_T=\varepsilon_{ij}n^\prime_in^\prime_j$, 
where ${\bf n}$ and ${\bf n}^\prime$ are unit vectors in the original and transverse 
directions, respectively.

The number of independent elastic constants is determined by the dimensionality and 
rotational symmetry of the lattice in question.  For example, in two dimensions, square 
lattices have three independent elastic constants, and triangular lattices are ``elastically 
isotropic" (i.e., elastic properties are independent of direction) and thus have only two \cite{LandauLifshitz}.  
Lattices with cubic symmetry have three independent elastic constants, the least of any 
crystal structure in three dimensions.\\

\noindent {\it Two Dimensions: Triangular Lattice Example}\\

This analysis can be applied to any structure in two dimensions, but perhaps the 
simplest case is that of the triangular lattice.
The free energy (neglecting $g_0$, an irrelevant offset), is 
\begin{equation}
g_{tri}=v\left\{2\lambda_1(\varepsilon_{xx}+\varepsilon_{yy})^2+
\lambda_2\left[(\varepsilon_{xx}-\varepsilon_{yy})^2 + 4\varepsilon_{xy}^2\right]\right\},
\label{free-energy-triangular}
\end{equation}	
where $\lambda_1$ and $\lambda_2$ are the elastic constants, and $v$ 
is the two-dimensional ``volume," namely, the area.  Since both terms are quadratic 
invariants of the strain tensor, $g$ is rotationally invariant.  This property holds true 
of the Poisson's ratio, which is
\begin{equation}
\nu_{tri} = \frac{2\lambda_1-\lambda_2}{2\lambda_1+\lambda_2}.\label{poissons-ratio-triangular}
\end{equation}
The free energy expansion must be a positive-definite quadratic form, or else the lattice is 
unstable.  Thus, the eigenvalues of the Hessian matrix of $g_{tri}$ 
must be positive. This is true if and only if $\lambda_1>0$ and $\lambda_2>0$.  The Poisson's ratio must thus fall within the range 
$-1<\nu_{tri}<1$.  By expanding Eq. (\ref{energy}) as well as the area to 
quadratic order in the strain tensor, we obtain the elastic constants in 
terms of the first and second derivatives of the pair potential, evaluated at the 
neighbor distances of the lattice:
\begin{eqnarray}
& \lambda_1 = \frac{1}{32v} \sum_{i=1}^N \left[ |{\bf r}_i |^2 \phi^{\prime\prime} ( |{\bf r}_i | ) 
- |{\bf r}_i | \phi^\prime ( | {\bf r}_i |) \right], \nonumber \\
& \lambda_2= \frac{1}{4v} \sum_{i=1}^N \left[   \left( \frac{x_i^2 y_i^2}{|{\bf r}_i |^2}\right)  
\phi^{\prime\prime} ( |{\bf r}_i | ) + \left(  \frac{|{\bf r}_i |}{2}- \frac{x_i^2 y_i^2}{|{\bf r}_i |
^3}\right) \phi^\prime ( | {\bf r}_i |) \right],
\end{eqnarray}     
where the sum is over every particle except that at the 
origin, and $x_i$ and $y_i$'s are the Cartesian coordinates of the particles in the 
unstrained lattice.  A straightforward calculation shows 
that the Poisson's ratio, given in Eq. (\ref{poissons-ratio-triangular}), can be written as
\begin{equation}
\nu_{tri} = \frac{1+2p\kappa_T}{3-2p\kappa_T},\label{poissons-ratio-triangular-2}
\end{equation}  
where $\kappa_T$ and $p$ are defined by:
\begin{eqnarray}
& \frac{1}{\kappa_T} = \left[ -v \frac{dp}{dv}\right] =  \frac{1}{8v}\sum_{i=1}^N \left[ |{\bf r}_i |
^2 \phi^{\prime\prime} ( |{\bf r}_i | ) - |{\bf r}_i | \phi^\prime ( | {\bf r}_i |) \right], \mbox{ and}\\ 
\nonumber
& p = -\frac{du}{dv}= -\frac{1}{4v}\sum_{i=1}^N |{\bf r}_i | \phi^\prime ( | {\bf r}_i |).
\end{eqnarray}
Thus, the Poisson's ratio, given in Eq. (\ref{poissons-ratio-triangular-2}), cannot 
be negative as long as $p>0$, since for mechanical stability, $
\kappa_T>0$.  Importantly, it is only for negative pressures, i.e., when the system is under 
tension, that $\nu_{tri}$ can be negative.

\begin{figure}
\includegraphics[width=2.4in]{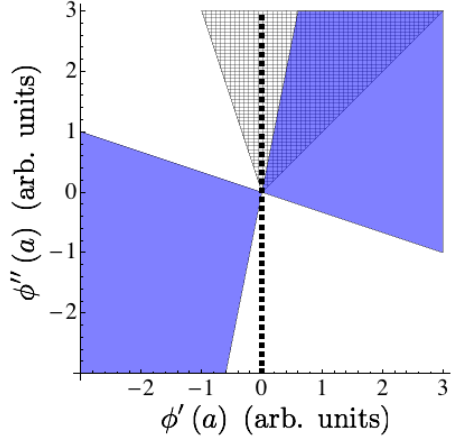}
\caption{(Color online) Regime of negative Poisson's ratio in a triangular lattice, where 
the pair potential extends to the first neighbor only.  The parameter space is composed of 
the first and second derivatives of the pair potential evaluated at the nearest-neighbor 
distance.  In the continuously shaded region,  $\nu_{tri}$, given in Eq. (\ref{poissons-ratio-triangular}), 
is negative.  In the grid-shaded region, the lattice is stable.  The overlap 
region is given in Eq. (\ref{triangular-inequalities}).  The pressure is positive to the left of 
the dotted line and negative to the right.    }       
\label{triangular-regionplot}
\end{figure}

Even if the interaction potential extends only to the nearest neighbor in the triangular 
lattice and is zero beyond it, NPR behavior can be achieved at negative 
pressure, as long as:
\begin{equation}
\phi '(a)>0,\mbox{ }\phi '(a)<a \phi ''(a)<5 \phi '(a)\label{triangular-inequalities}
\end{equation}  
where $a=\sqrt{  2v / \sqrt{3} }$ is the lattice constant.  Figure \ref{triangular-regionplot} 
depicts these inequalities in a parameter space of $\phi^\prime(a)$ and $\phi^{\prime
\prime}(a)$. 
In the case that the potential extends beyond the first neighbor, corresponding inequalities to those in Eq. (\ref{triangular-inequalities}) 
can be found, involving the first and second derivatives of the potential at relevant neighbor distances.

To understand intuitively why NPR behavior only occurs at negative 
pressure, consider the function $\Phi(r^2)=\phi(r)$, which is the 
interaction potential rewritten in terms of the square of the inter-particle distance.  
A condition for lattice stability is $\Phi^{\prime\prime}(a^2)>0$, which 
physically means that the ``effective repulsive spring constant" between nearest-neighbor 
particles must decrease with distance.  Thus, if the bonds 
between neighboring particles are thought of as springs with zero rest length, an imposed 
outward strain has the effect of weakening the effective spring constants of the 
bonds in the transverse direction.  Under positive pressure, such a 
weakening causes a contraction, but under tension it causes an expansion.          

\begin{figure}
\includegraphics[width=2.7in]{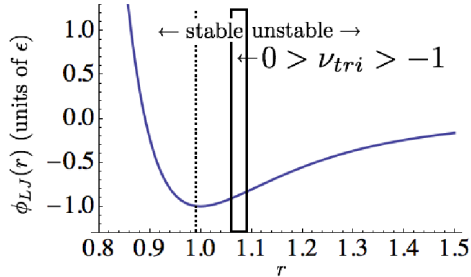}
\caption{(Color online) Region of lattice constants (indicated by the rectangular box) for 
which the Poisson's ratio is negative in a triangular lattice, using the LJ 
interaction potential, $\phi_{LJ}$.  Pressure is positive to the left of the dotted line and 
negative to the right; thus, NPR behavior only occurs at negative pressure.  To the right of 
the rectangular box, the lattice becomes unstable. }       
\label{lennard-jones-plot}
\end{figure}

When $\phi$ is taken to be the well-known 12-6 Lennard-Jones (LJ) potential, 
given by $\phi_{LJ}(r)=\epsilon\left[(b/r)^{12}-2(b/r)^6\right]$, there is a range of lattice constants for which NPR behavior is exhibited, namely $1.0596b < a < 
1.0870b$, as demonstrated in Fig. \ref{lennard-jones-plot}.  This corresponds to a 
pressure range of $-4.6363(\epsilon/b^2)>p>-4.8516(\epsilon/b^2)$.  Therein, the 
Poisson's ratio takes on values of 0 through $-1$, the lower bound for stability.  Note that a sufficient number of neighboring particles are included to 
accurately calculate the energy.  That the Poisson's ratio is 
negative in this system is a surprising result, since the simple LJ 
potential has been studied extensively, and was not thought to exhibit this behavior.    

Other two-dimensional crystals have been studied for a general $\phi$, including the 
square lattice and non-Bravais crystals such as the honeycomb and kagom\'e.  In all 
cases, NPR behavior was only found under tension.         \\

\noindent {\it Three Dimensions: Lattices with Cubic Symmetry}\\

We proceed here with a calculation for lattices with cubic symmetry that is analogous to 
that for the triangular lattice.  The most common examples of these are 
the face-centered cubic (fcc), the body-centered cubic (bcc) and the simple cubic (sc) 
lattices.  Such systems have three independent elastic constants and are generally not 
elastically isotropic.  The most general expression for the Gibbs free energy of 
cubic systems is,
\begin{equation}
g = v\left\{\frac{1}{2}\lambda_1(\varepsilon_{xx}^2+\varepsilon_{yy}^2+
\varepsilon_{zz}^2)
+\lambda_2(\varepsilon_{xx}\varepsilon_{yy}+\varepsilon_{xx}\varepsilon_{zz}+
\varepsilon_{yy}\varepsilon_{zz})+2\lambda_3(\varepsilon_{xy}^2+\varepsilon_{xz}^2+
\varepsilon_{yz}^2)  \right\} ,\label{free-energy-cubic}
\end{equation} 
where $\lambda_1,\lambda_2,$ and $\lambda_3$ are the elastic constants, and $v$ is 
the specific volume.  The Poisson's ratio is a function of both the direction of the imposed 
strain and the chosen transverse direction.  As an example, if the strain is imposed along 
the [100] direction, the Poisson's ratio is
\begin{equation}
\nu_{cubic}^{100} = \frac{\lambda_2}{\lambda_1+\lambda_2}.  \label{poissons-ratio-cubic}
\end{equation}    
In order for the Hessian of the quadratic form given in Eq. (\ref{free-energy-cubic}) to be 
positive definite, a necessary condition for lattice stability, we must have that $
\lambda_1>0$, $\lambda_3>0$ and $-\lambda_1/2<\lambda_2<\lambda_1$.  If we 
expand Eq. (\ref{energy}) to quadratic order in the strain tensor components, we obtain
\begin{eqnarray}
& \lambda_1 = -p+\frac{1}{2v} \sum_{i=1}^N \left(    \frac{x_i}{|{\bf r}_i|}    \right)^4\left[ |{\bf 
r}_i |^2 \phi^{\prime\prime} ( |{\bf r}_i | ) - |{\bf r}_i | \phi^\prime ( | {\bf r}_i |) \right], 
\nonumber \\
& \lambda_2= \frac{p}{2}+\frac{3}{2\kappa_T}    -   \frac{1}{4v} \sum_{i=1}^N 
\left(    \frac{x_i}{|{\bf r}_i|}    \right)^4\left[ |{\bf r}_i |^2 \phi^{\prime\prime} ( |{\bf r}_i | ) - |{\bf 
r}_i | \phi^\prime ( | {\bf r}_i |) \right], \nonumber \\
& \lambda_3= -p   +   \frac{1}{2v} \sum_{i=1}^N \left(    \frac{x_i y_i}{|{\bf r}_i|^2}    
\right)^2\left[ |{\bf r}_i |^2 \phi^{\prime\prime} ( |{\bf r}_i | ) - |{\bf r}_i | \phi^\prime ( | {\bf r}_i |) 
\right], \mbox{ with} \nonumber \\
& \frac{1}{\kappa_T}= \frac{1}{18v} \sum_{i=1}^N \left(    \frac{x_i}{|{\bf r}_i|^2}    
\right)^2\left[ |{\bf r}_i |^2 \phi^{\prime\prime} ( |{\bf r}_i | ) - |{\bf r}_i | \phi^\prime ( | {\bf r}_i |) 
\right], \mbox{ and}\nonumber \\ 
& p=-\frac{1}{6v}\sum_{i=1}^N |{\bf r}_i | \phi^\prime ( | {\bf r}_i |).\label{elastic-constants-cubic}
\end{eqnarray}     
where $p$ is the pressure and $\kappa_T$ is the compressibility.  Note that these 
expressions apply equally well to any Bravais lattice with cubic symmetry.  If we impose 
an additional linear constraint on the elastic constants, we find that we can impose elastic 
isotropy on the system, namely, if we enforce the following:
\begin{equation}
\lambda_1=\lambda_2+2\lambda_3,\label{isotropic-cubic}
\end{equation}  
then the system becomes elastically isotropic because the free energy given in Eq. 
(\ref{free-energy-cubic}) can be written as a function only of quadratic invariants of the 
strain tensor.  Even with this additional applied constraint, we find that the system can 
exhibit NPR behavior under tension, with the Poisson's ratio in any direction given by Eq. 
(\ref{poissons-ratio-cubic}).  The Poisson's ratio must fall between $-1$ and $+1/2$ in this 
case.  If we consider a particular lattice we can, by employing Eqs. (\ref{poissons-ratio-cubic}), 
(\ref{elastic-constants-cubic}) and (\ref{isotropic-cubic}) find inequalities, which 
involve the pair potential evaluated at the neighbor distances, that describe the regime in 
which the Poisson's ratio is negative and the system is elastically isotropic.  For each set 
of coordination shells included in the calculation, such inequalities can be found; none 
were found to allow for NPR behavior at positive pressure.

Consider as an example the fcc lattice, which has 12 nearest neighbors (at distance $a/
\sqrt{2}$) and 6 next-nearest neighbors at distance $a$, where $a$ is the side length of 
the cubic cell.  If the potential extends only to the nearest neighbor, NPR behavior is not 
possible.  Both NPR behavior and elastic isotropy can be exhibited at negative pressure if 
the pair potential extends to the nearest and next-nearest neighbors, if the following 
constraints are satisfied:
\begin{eqnarray}
& 4 \phi '(a)-4 a \phi ''(a)+a \phi ''\left(\frac{a}{\sqrt{2}}\right)=\sqrt{2} \phi'\left(\frac{a}
{\sqrt{2}}\right),\nonumber \\
& \mbox{ } \sqrt{2} \phi '\left(\frac{a}{\sqrt{2}}\right)<a \phi ''\left(\frac{a}{\sqrt{2}}\right)<4 \phi 
'(a)+5 \sqrt{2} \phi '\left(\frac{a}{\sqrt{2}} \right) ,\mbox{ and }\nonumber \\
&  4 \phi '(a)+9 \sqrt{2}
   \phi '\left(\frac{a}{\sqrt{2}}\right)<5 a \phi ''\left(\frac{a}{\sqrt{2}}\right).
\end{eqnarray}    	
At zero pressure, the Poisson's ratio goes to $1/4$, as predicted 
by the Cauchy relations \cite{CauchyRelations}.    

Our analysis suggests that NPR behavior does not occur at positive pressures in crystals 
when the system contains only pair interactions, and the material is elastically isotropic.  
However, we present here a three-body potential that yields NPR behavior in close-
packed two and three dimensional lattices by construction at zero temperature and 
positive pressure.  In order to produce this behavior, the potential has a built-in energy 
cost associated with deforming the equilateral triangles in the two-dimensional triangular 
lattice and the three-dimensional close-packed lattices.  The three-body potential is given 
by
\begin{equation}
\phi_3(r,s,t)=\alpha f(r) f(s) f(t) F(r, s, t),
\end{equation}
where $\alpha$ is a positive constant, $r$, $s$, and $t$ are the side lengths of a triangle 
defined by a triplet of particles, $f$ is some function that goes to zero sufficiently quickly 
that only nearest neighbors are within range of the potential, and 
\begin{equation}
F(r,s,t)=  \frac{(r+s+t)^2}{3^{3/2}\left[   2\left(   r^2s^2+r^2t^2+s^2t^2  \right) -r^4 - s^4 - t^4  
\right]^{1/2}}-1,
\end{equation}
a function that is zero if the triplet of particles defines an equilateral triangle, but is greater 
than zero otherwise.  Thus, if an outward strain is imposed on the system, then regardless 
of the ambient pressure, the system will expand in the transverse direction, if $\alpha$ is 
sufficiently large.

In conclusion, we have shown that in two and three dimensions, classical systems with only pair potentials can have a negative Poisson's ratio at zero 
temperature, a surprising result.  To the authors' knowledge, this has not previously been made explicit \cite{Footnote}.  However, this auxetic behavior is only present when the system is at 
negative pressure, and thus not in thermal equilibrium.  NPR 
materials may potentially be experimentally produced using synthetic techniques that rely 
on kinetic effects; examples include tempered glass \cite{TemperedGlass}, and even 
colloidal crystals \cite{ColloidalCrystalsUnderTension}.  In two dimensions, it was proved that NPR behavior could be 
found at negative pressure; the proof was shown here for the triangular lattice, 
but a similar result also holds true for the square lattice.  In three dimensions, a set of 
constraints on the pair interaction was found such that, if satisfied, the fcc lattice is both 
elastically isotropic and has NPR behavior at negative pressure.  Although the fcc was 
chosen as an example, the calculation may be generalized given the 
expressions for the elastic constants of cubic systems reported here.  We also presented 
a three-body interaction potential that by construction gives rise to a solid with elastic 
isotropy and NPR behavior at zero temperature and arbitrary pressure (negative and 
positive values).  This suggests that the requirement of negative pressure is limited to 
systems with only pair interactions.  A general proof of such a statement does not exist 
and will be considered in future work.  In other future work, we hope to describe under 
what conditions NPR behavior can be observed in colloidal crystals under tension, given 
the experimentally realizable interaction potentials between colloidal particles.  Finding 
NPR behavior over a wide range in temperature and pressure is a challenging 
\emph{optimization} problem that we also intend to address. 

M.C.R. acknowledges the support of the Natural Sciences and Engineering Research 
Council of Canada.  S. T. thanks the Institute for Advanced Study for their hospitality 
during his stay there.  This work was supported by the Office of Basic Energy Sciences, 
DOE, under Grant No. 
DE-FG02-04ER46108.


\begin{thebibliography}{17}
\expandafter\ifx\csname natexlab\endcsname\relax\def\natexlab#1{#1}\fi
\expandafter\ifx\csname bibnamefont\endcsname\relax
  \def\bibnamefont#1{#1}\fi
\expandafter\ifx\csname bibfnamefont\endcsname\relax
  \def\bibfnamefont#1{#1}\fi
\expandafter\ifx\csname citenamefont\endcsname\relax
  \def\citenamefont#1{#1}\fi
\expandafter\ifx\csname url\endcsname\relax
  \def\url#1{\texttt{#1}}\fi
\expandafter\ifx\csname urlprefix\endcsname\relax\def\urlprefix{URL }\fi
\providecommand{\bibinfo}[2]{#2}
\providecommand{\eprint}[2][]{\url{#2}}

\bibitem{paper149}  
O. Sigmund, S. Torquato and I.A. Aksay, J. Mater. Res. {\bf 14}, 1038 (1998).

\bibitem[{\citenamefont{Xu et~al.}(1999)\citenamefont{Xu, Arias, Brittain,
  Zhao, Grzybowski, Torquato, and Whitesides}}]{TorquatoWhitesidesNPR}
\bibinfo{author}{\bibfnamefont{B.}~\bibnamefont{Xu}},
  \bibinfo{author}{\bibfnamefont{F.}~\bibnamefont{Arias}},
  \bibinfo{author}{\bibfnamefont{S.~T.} \bibnamefont{Brittain}},
  \bibinfo{author}{\bibfnamefont{X.~M.} \bibnamefont{Zhao}},
  \bibinfo{author}{\bibfnamefont{B.}~\bibnamefont{Grzybowski}},
  \bibinfo{author}{\bibfnamefont{S.}~\bibnamefont{Torquato}}, \bibnamefont{and}
  \bibinfo{author}{\bibfnamefont{G.~M.} \bibnamefont{Whitesides}},
  \bibinfo{journal}{Adv. Mat.} \textbf{\bibinfo{volume}{11}},
  \bibinfo{pages}{1186} (\bibinfo{year}{1999}).


\bibitem[{\citenamefont{Choi and Lakes}(1991)}]{ChoiFasteners}
\bibinfo{author}{\bibfnamefont{J.~B.} \bibnamefont{Choi}} \bibnamefont{and}
  \bibinfo{author}{\bibfnamefont{R.~S.} \bibnamefont{Lakes}},
  \bibinfo{journal}{Cell. Polym.} \textbf{\bibinfo{volume}{10}},
  \bibinfo{pages}{205} (\bibinfo{year}{1991}).


\bibitem[{\citenamefont{Lakes}(1987)}]{Lakes1987}
\bibinfo{author}{\bibfnamefont{R.}~\bibnamefont{Lakes}},
  \bibinfo{journal}{Science} \textbf{\bibinfo{volume}{235}},
  \bibinfo{pages}{1038} (\bibinfo{year}{1987}).

\bibitem[{\citenamefont{Milton}(1992)}]{MiltonComposites}
\bibinfo{author}{\bibfnamefont{G.~W.} \bibnamefont{Milton}},
  \bibinfo{journal}{J. Mech. Phys. Solids} \textbf{\bibinfo{volume}{40}},
  \bibinfo{pages}{1105} (\bibinfo{year}{1992}).

\bibitem[{\citenamefont{Friis et~al.}(1988)\citenamefont{Friis, Lakes, and
  Park}}]{MetallicPolymericFoams}
\bibinfo{author}{\bibfnamefont{E.~A.} \bibnamefont{Friis}},
  \bibinfo{author}{\bibfnamefont{R.~S.} \bibnamefont{Lakes}}, \bibnamefont{and}
  \bibinfo{author}{\bibfnamefont{J.~B.} \bibnamefont{Park}},
  \bibinfo{journal}{J. Mat. Sci.} \textbf{\bibinfo{volume}{23}},
  \bibinfo{pages}{4406} (\bibinfo{year}{1988}).

\bibitem[{\citenamefont{Almgren}(1985)}]{AlmgrenHinges}
\bibinfo{author}{\bibfnamefont{R.~F.} \bibnamefont{Almgren}},
  \bibinfo{journal}{J. Elasticity} \textbf{\bibinfo{volume}{15}},
  \bibinfo{pages}{427} (\bibinfo{year}{1985}).

\bibitem[{\citenamefont{Baughman et~al.}(1998)\citenamefont{Baughman,
  Shacklette, Zakhidov, and Stafstrom}}]{BaughmanNPR}
\bibinfo{author}{\bibfnamefont{R.~H.} \bibnamefont{Baughman}},
  \bibinfo{author}{\bibfnamefont{J.~M.} \bibnamefont{Shacklette}},
  \bibinfo{author}{\bibfnamefont{A.~A.} \bibnamefont{Zakhidov}},
  \bibnamefont{and}
  \bibinfo{author}{\bibfnamefont{S.}~\bibnamefont{Stafstrom}},
  \bibinfo{journal}{Nature} \textbf{\bibinfo{volume}{392}},
  \bibinfo{pages}{362} (\bibinfo{year}{1998}).

\bibitem[{\citenamefont{Rechtsman
  et~al.}(2007{\natexlab{a}})\citenamefont{Rechtsman, Stillinger, and
  Torquato}}]{RechtsmanNTE}
\bibinfo{author}{\bibfnamefont{M.~C.} \bibnamefont{Rechtsman}},
  \bibinfo{author}{\bibfnamefont{F.~H.} \bibnamefont{Stillinger}},
  \bibnamefont{and} \bibinfo{author}{\bibfnamefont{S.}~\bibnamefont{Torquato}},
  \bibinfo{journal}{J. Phys. Chem. A} \textbf{\bibinfo{volume}{111}},
  \bibinfo{pages}{12816} (\bibinfo{year}{2007}{\natexlab{a}}).

\bibitem[{\citenamefont{Rechtsman et~al.}(2005)\citenamefont{Rechtsman,
  Stillinger, and Torquato}}]{RechtsmanPRL}
\bibinfo{author}{\bibfnamefont{M.~C.} \bibnamefont{Rechtsman}},
  \bibinfo{author}{\bibfnamefont{F.~H.} \bibnamefont{Stillinger}},
  \bibnamefont{and} \bibinfo{author}{\bibfnamefont{S.}~\bibnamefont{Torquato}},
  \bibinfo{journal}{Phys. Rev. Lett.} \textbf{\bibinfo{volume}{95}},
  \bibinfo{pages}{228301} (\bibinfo{year}{2005});
\bibinfo{author}{\bibfnamefont{M.~C.} \bibnamefont{Rechtsman}},
  \bibinfo{author}{\bibfnamefont{F.~H.} \bibnamefont{Stillinger}},
  \bibnamefont{and} \bibinfo{author}{\bibfnamefont{S.}~\bibnamefont{Torquato}},
  \bibinfo{journal}{Phys. Rev. E} \textbf{\bibinfo{volume}{74}},
  \bibinfo{pages}{21404} (\bibinfo{year}{2006});
\bibinfo{author}{\bibfnamefont{M.~C.} \bibnamefont{Rechtsman}},
  \bibinfo{author}{\bibfnamefont{F.~H.} \bibnamefont{Stillinger}},
  \bibnamefont{and} \bibinfo{author}{\bibfnamefont{S.}~\bibnamefont{Torquato}},
  \bibinfo{journal}{Phys. Rev. E} \textbf{\bibinfo{volume}{75}},
  \bibinfo{pages}{31403} (\bibinfo{year}{2007}{\natexlab{b}}).

\bibitem[{\citenamefont{Landau and Lifshitz}(1986)}]{LandauLifshitz}
\bibinfo{author}{\bibfnamefont{E.~D.} \bibnamefont{Landau}} \bibnamefont{and}
  \bibinfo{author}{\bibfnamefont{E.~M.} \bibnamefont{Lifshitz}},
  \emph{\bibinfo{title}{{Theory of Elasticity}}} (\bibinfo{publisher}{Pergamon
  Press}, \bibinfo{address}{Oxford, UK}, \bibinfo{year}{1986}).

\bibitem[{\citenamefont{Seitz}(1940)}]{CauchyRelations}
\bibinfo{author}{\bibfnamefont{F.}~\bibnamefont{Seitz}},
  \emph{\bibinfo{title}{{The Modern Theory of Solids}}}
  (\bibinfo{publisher}{McGraw-Hill}, \bibinfo{address}{Ann Arbor, MI},
  \bibinfo{year}{1940}).

\bibitem{Footnote}
NPR behavior was  found in network systems (vertices
connected by flexible bonds) under tension
\cite{BoalNetworks}.
 The present work is fundamentally different in that we are considering a many-particle system in which an arbitrary number of neighboring particles may be incorporated and the pair potential is completely general. Moreover, unlike Ref. \cite{BoalNetworks}, we do not use a mean-field approximation.

\bibitem[{\citenamefont{Carr{\'e} and Daudeville}(1999)}]{TemperedGlass}
\bibinfo{author}{\bibfnamefont{H.}~\bibnamefont{Carr{\'e}}} \bibnamefont{and}
  \bibinfo{author}{\bibfnamefont{L.}~\bibnamefont{Daudeville}},
  \bibinfo{journal}{J. Eng. Mech.} \textbf{\bibinfo{volume}{125}},
  \bibinfo{pages}{914} (\bibinfo{year}{1999}).

\bibitem[{\citenamefont{Zhang et~al.}(2008)\citenamefont{Zhang, Liu, Wang, and
  Ming}}]{ColloidalCrystalsUnderTension}
\bibinfo{author}{\bibfnamefont{J.}~\bibnamefont{Zhang}},
  \bibinfo{author}{\bibfnamefont{H.}~\bibnamefont{Liu}},
  \bibinfo{author}{\bibfnamefont{Z.}~\bibnamefont{Wang}}, \bibnamefont{and}
  \bibinfo{author}{\bibfnamefont{N.}~\bibnamefont{Ming}}, \bibinfo{journal}{J.
  Appl. Phys.} \textbf{\bibinfo{volume}{103}}, \bibinfo{eid}{013517}
  (\bibinfo{year}{2008}).

\bibitem[{\citenamefont{Boal et~al.}(1993)\citenamefont{Boal, Seifert, and
  Shillcock}}]{BoalNetworks}
\bibinfo{author}{\bibfnamefont{D.~H.} \bibnamefont{Boal}},
  \bibinfo{author}{\bibfnamefont{U.}~\bibnamefont{Seifert}}, \bibnamefont{and}
  \bibinfo{author}{\bibfnamefont{J.~C.} \bibnamefont{Shillcock}},
  \bibinfo{journal}{Phys. Rev. E} \textbf{\bibinfo{volume}{48}},
  \bibinfo{pages}{4274} (\bibinfo{year}{1993}).



\end{thebibliography}
\end{document}